# Lanthanide substitution effects in electron-doped high-$T_\text{c}$ superconductors studied by angle-resolved photoemission spectroscopy

M. Ikeda[1], T. Yoshida[1], A. Fujimori[1],

M. Kubota[2], K. Ono[2], K. Unozawa[3], T. Sasagawa[3], and H. Takagi[3]

[1]*Department of Complexity Science and Engineering,*

*University of Tokyo, Kashiwanoha 5-1-5, Kashiwa, Chiba 277-8561, Japan*

[2]*Institute of Material Structures Science, High Energy Accelerator Research Organization (KEK),*

*Oho 1-1, Tsukuba, Ibaraki 305-0801, Japan*

[3]*Department of Advanced Materials Science, University of Tokyo,*

*Kashiwanoha 5-1-5, Kashiwa, Chiba 277-8561, Japan*

## ABSTRACT

We have performed an angle-resolved photoemission study of the electron-doped high-$T_\text{c}$ superconductors (HTSCs) $Sm_{1.85}Ce_{0.15}CuO_4$ (SCCO) and $Eu_{1.85}Ce_{0.15}CuO_4$ (ECCO). Around the nodal point $\vec{k} \sim (\pi/2, \pi/2)$, we observed an energy gap at the Fermi level ($E_\text{F}$) for both samples, resulting from the strong effects of antiferromagnetism. The magnitude of this gap in ECCO is larger than that in SCCO, implying that the effects of antiferromagnetism in ECCO are stronger than that in SCCO. Also, we quantitatively confirmed the effects of antiferromagnetism by the band dispersions to tight-binding calculation.

## 1. INTRODUCTION

Since the discovery of electron-doped high-$T_c$ superconductors (HTSCs) [1], a number of experiments and theoretical studies have been performed to elucidate the similarity and dissimilarity between the electron-doped HTSCs and hole-doped ones. In addition to the comparison between them, the electron-doped HTSCs possess interesting physical properties in its own right. The electron doped HTSCs $Ln_{2-x}Ce_xCuO_4$ ($Ln$ = La, Pr, Nd, Sm, Eu) with the $T'$-type structure exhibit systematic changes in various physical properties as a function of the ionic radius of $Ln^{3+}$. As the ionic radius of $Ln^{3+}$ becomes smaller, $T_c$ becomes smaller [2, 3] and the antiferromagnetic phase is extended to higher doping levels in the $x$-$T$ phase diagram [4, 5]. Above all, $Sm_{2-x}Ce_xCuO_4$ shows the remarkable property that the superconducting phase is enclosed by the antiferromagnetic phase, namely, exists within the wide antiferromagnetic phase in the $x$-$T$ phase diagram. In case of ECCO, the superconducting phase disappears in the $x$-$T$ phase diagram. Therefore, in order to elucidate the interplay between the superconductivity and the antiferromagnetism, it has been highly desired to perform angle-resolved photoemission spectroscopy (ARPES) studies of these systems.

## 2. EXPERIMENT

High-quality single crystals of optimally doped $Sm_{1.85}Ce_{0.15}CuO_4$ (SCCO) and $Eu_{1.85}Ce_{0.15}CuO_4$ (ECCO) were grown by the traveling solvent floating zone method. The single crystals of SCCO and ECCO were annealed at 900°C for 24 hours under Ar flow to remove excess oxygen. The $T_c$ of

SCCO was ~16 K and ECCO did not show superconductivity.

The ARPES measurements were performed at beamline 28A of Photon Factory (PF), High Energy Accelerators Research Organization (KEK), using circularly polarized photons with energy of 55 eV. The incident photon beam angle was at approximately 45° to the sample surface. We used a SCIENTA SES-2002 electron-energy analyzer in the angle mode, where one can collect spectra over ~14°, corresponding to a momentum width of ~1.1π (in units of 1/a, where a = 3.9 Å is the lattice constant). The total energy resolution and angular (momentum) resolution were 15 meV and 0.2° (0.01π), respectively. Samples were cleaved *in situ* under an ultrahigh vacuum of $10^{-9}$ Pa to obtain clean surfaces and measured at ~10 K. We used a five axes manipulator [6] and the sample temperature was measured by a calibrated diode mounted near the sample. The evaporated gold was used to determine the Fermi level ($E_F$) of the samples. The spectral intensity was normalized to the intensity above $E_F$, which arose from the second order light of the monochromator.

## 3. RESULTS AND DISCUSSION

Figure 1(a) and (b) show the plot of ARPES intensity near the $E_F$ in SCCO and ECCO, respectively, as a function of two-dimensional wave vector. The energy distribution curves (EDCs) have been integrated within a ±30 meV window of $E_F$. In Fig. 1, the Fermi surfaces calculated by the tight-binding model (see below) are also depicted. On the calculated Fermi surface, there is a region in which the spectral weight is suppressed, which is consistent with the previous ARPES studies of $Nd_{2-x}Ce_xCuO_4$ [7, 8]. These regions are called ``hot spot'', which is caused by (π, π) scattering due to the effects of antiferromagnetism. From comparison between Fig. 1(a) and (b), the intensity near ($\pi/2$, $\pi/2$) in ECCO is weak compared with that in SCCO. This suggests that the effects of antiferromagnetism are stronger in ECCO than in SCCO.

In order to examine the effects of antiferromagnetism, we show ARPES intensity plots near the nodal direction for SCCO and ECCO in Fig. 2. The directions of the cuts are depicted in each inset. As shown in Fig. 2, the energy gap in the nodal direction was observed for the both samples. The size of the gap of ECCO is larger than that of SCCO, indicating the stronger effects of antiferromagnetism in ECCO than in SCCO. Since the antiferromagnetic state tends to suppress superconductivity, the present result explains the absence of $T_c$ in ECCO.

Next, in order to quantitatively evaluate the effects of antiferromagnetism from the ARPES spectra, we employed the model with antiferromagnetic shadow bands as in Ref. [8]. Here, we have used a two-dimensional tight-binding (TB) model including the effects of antiferromagnetism. The TB dispersion is given by

$$E = \varepsilon_0 \pm \sqrt{\Delta^2 + 4t^2 \left(\cos k_x a + \cos k_y a\right)^2} - 4t' \cos k_x a \cos k_y a ,$$

where $t$, $t'$, $\varepsilon_0$, and $\Delta$ represent transfer integrals between the nearest and next-nearest-neighbor Cu sites, the center of the energy band and the energy difference between the spin-up and spin-down sites, respectively. In the TB calculation, first, in order to obtain the values of $t'/t$ and $\varepsilon_0/t$, we fitted the peak positions of the momentum distribution curves (MDCs) near $E_F$ to the TB calculation with $\Delta = 0$ shown in Fig. 1. For SCCO, the TB Fermi surface $t'/t = -0.23\pm0.05$ and $\varepsilon_0/t = 0.26\pm0.08$ well correspond to the intensity distribution at $E_F$, while for ECCO the TB Fermi surface $t'/t = -0.21\pm$

0.05 and $\varepsilon_0/t = 0.25\pm0.07$ well correspond to the intensity distribution at $E_F$. Secondly, by using the values of $t'/t$ and $\varepsilon_0/t$ thus obtained, the MDC peak positions in the nodal direction were fitted to the TB band dispersion to obtain the parameters $t$ and $\Delta$. The fitted value of $\Delta$ is $0.09\pm0.03$ eV and $0.13\pm0.03$ eV for SCCO and ECCO, respectively. The energy difference of $\Delta$ between them ~0.03 eV corresponds to the difference in the nodal gap as shown in Fig. 2. Fitted values of $t$ are $0.30\pm0.05$ eV and $0.36\pm0.05$ eV for SCCO and ECCO, respectively.

Finally, we discuss the reason why the effects of antiferromagnetism in ECCO are larger than that in SCCO. The effects of antiferromagnetism may be expressed by the superexchange interaction $J = \dfrac{4t_{pd}^4}{\Delta^2}\left(\dfrac{1}{\Delta} + \dfrac{1}{U}\right)$. It has been known that the larger the transfer integrals become, the larger $J$ becomes. Since the lattice constant of ECCO is smaller than that of SCCO, the $t$ of ECCO would be larger than that of SCCO, consistent with the present TB fitting results. Hence, the $J$ of ECCO would be larger than that of SCCO. Indeed, optical studies have shown that the value of $J$ in $Eu_2CuO_4$ is larger than that in $Sm_2CuO_4$ [9].

## 4. CONCLUSION

We performed an ARPES study of SCCO and ECCO in order to investigate the substitution effects of $Ln^{3+}$ in $Ln_{2-x}Ce_xCuO_4$. In the nodal direction, the energy gap of ECCO is larger than that of SCCO, indicating that the effects of antiferromagnetism in ECCO are strong compared with that in SCCO. Also, we quantitatively evaluated the effects of antiferromagnetism through the tight-binding model. The absence of $T_c$ in ECCO is explained by the strong effects of antiferromagnetism which opens an energy gap in the nodal direction. Such strong effects of antiferromagnetism in ECCO compared with SCCO may result from the difference in the lattice constants.

## 5. ACKNOWLEDGEMENT

We are grateful to Y. Aiura and N. Kamakura for technical support at beamline 28A of Photon Factory. This work was supported by a Grant-in-Aid for Scientific Research in Priority Area ``Invention of Anomalous Quantum Materials'' from the Ministry of Education, Culture, Sports, Science and Technology, Japan. This work was done under the approval of the Photon Factory Program Advisory Committee (Proposal No. 2006S2-001)

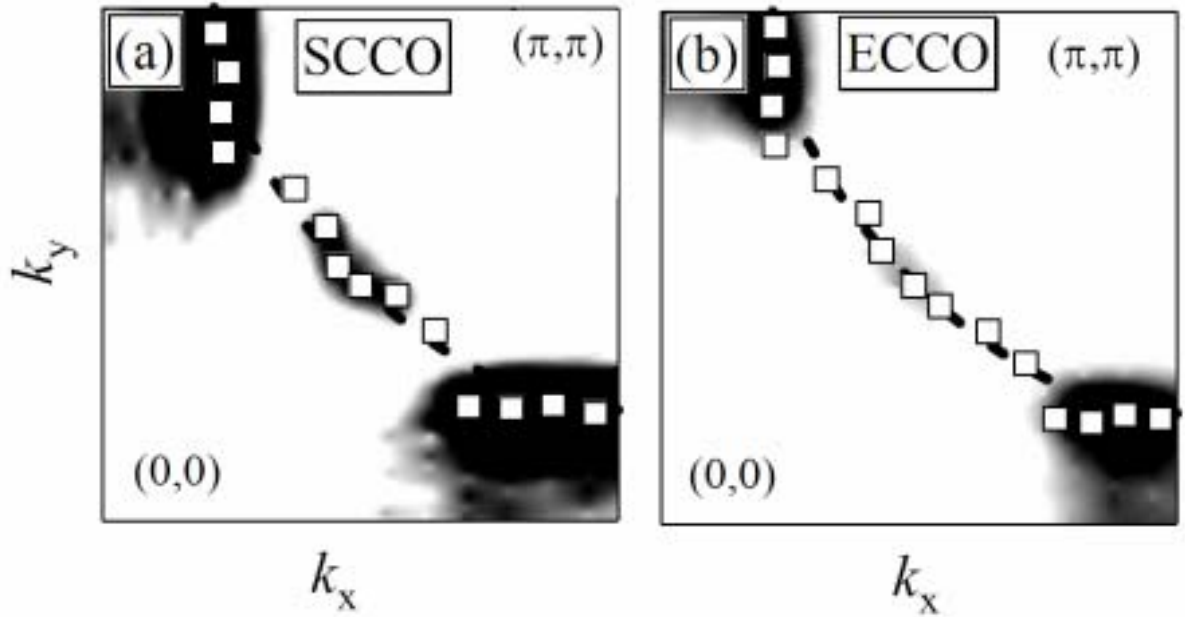

Figure 1 ARPES intensity near $E_F$ in (a) SCCO and (b) ECCO plotted as a function of the two-dimensional wave vector ($k_x$, $k_y$). Energy distribution curves have been integrated within a ±30 meV window around $E_F$. The data were taken over a Brillouin zone octant and symmetrized with respect to the (0, 0) - ($\pi$, $\pi$) line. Black dashed curves show the Fermi surfaces obtained by tight-binding calculation in the paramagnetic state. White squares show the MDC peak positions.

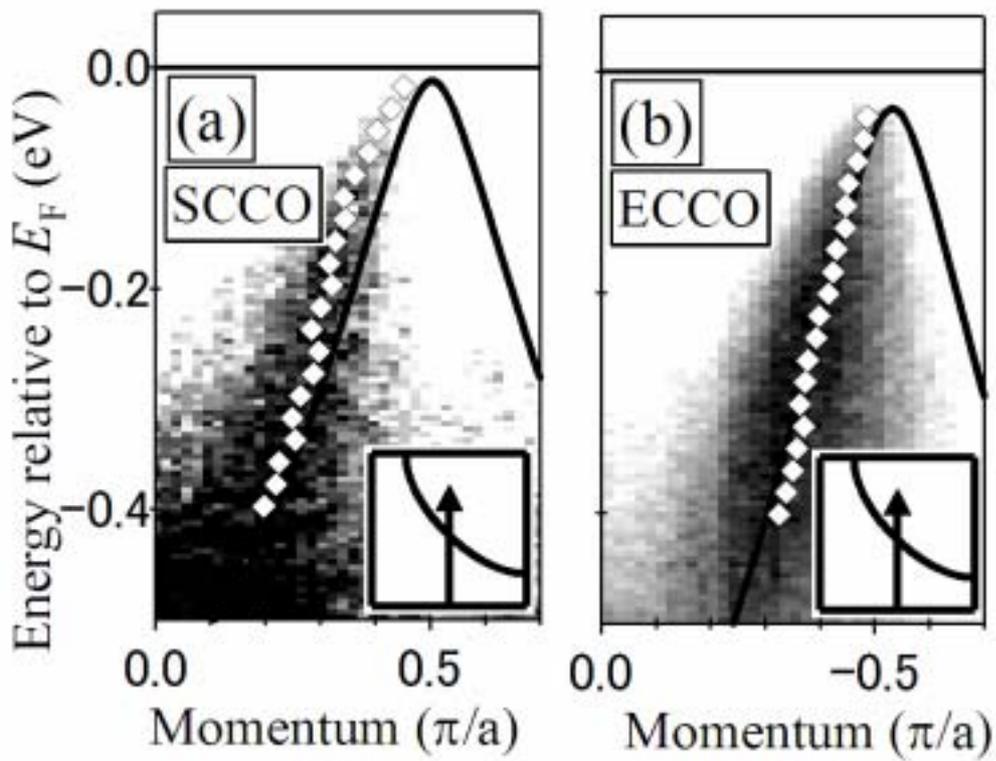

Figure 2 ARPES intensity plot along the cut including the nodal point for (a) SCCO and (b) ECCO. The direction of the cut is shown in the inset. White diamonds show momentum distribution curve (MDC) peak positions. Black curves are the results of tight-binding fitting in the antiferromagnetic state.